\documentclass[%
reprint,
 amsmath,amssymb,
 aps,
pre,
]{revtex4-1}

\usepackage{dcolumn}
\usepackage{bm}
\usepackage{graphicx}
\usepackage{subfigure}
\usepackage{color}
\usepackage{epsfig}
\usepackage{amssymb}
\usepackage{latexsym}

\newtheorem{prop}{Proposition}[section]

\newtheorem{cor}{Corollary}

\newtheorem{lm}{Lemma}

\newtheorem{thm}{Theorem}

\newcommand{\bthm}{\begin{thm}}
\newcommand{\ethm}{\end{thm}}

\newcommand{\bcor}{\begin{cor}}
\newcommand{\ecor}{\end{cor}}
\newcommand{\bprop}{\begin{prop}}
\newcommand{\eprop}{\end{prop}}
\newcommand{\blm}{\begin{lm}}
\newcommand{\elm}{\end{lm}}
\newcommand{\beq}{\begin{equation}}
\newcommand{\eeq}{\end{equation}}
\newcommand{\ber}{\begin{eqnarray}}
\newcommand{\eer}{\end{eqnarray}}

\newenvironment{proof1}{\begin{trivlist}\item[]{\bf Proof:\hspace{2mm}}}{\hfill$\blackbox$\end{trivlist}}

\newcommand{\blackbox}{\vrule height7pt width5pt depth1pt}

\newcommand{\bit}{\begin{itemize}}
\newcommand{\eit}{\end{itemize}}
\newcommand{\ben}{\begin{enumerate}}
\newcommand{\een}{\end{enumerate}}
\newcommand{\bdesc}{\begin{description}}
\newcommand{\edesc}{\end{description}}
\newcommand{\beqarrn}{\begin{eqnarray*}}
\newcommand{\eeqarrn}{\end{eqnarray*}}
\newenvironment{proofof}[1]{\begin{trivlist}\item[]{\bf Proof of #1:\hspace{2mm}
}}{\hfill\blackbox\end{trivlist}}
\newcommand{\bproofof}{\begin{proofof}}
\newcommand{\eproofof}{\end{proofof}}
\newenvironment{rem}{\begin{trivlist}\item[]{\bf
Remark:}\hspace{4mm}}{\end{trivlist}}
\newcommand{\brem}{\begin{rem}}
\newcommand{\erem}{\end{rem}}
\newenvironment{rems}{\begin{trivlist}\item[]{\bf
Remarks}\begin{itemize}}{\end{itemize}\end{trivlist}}
\newcommand{\brems}{\begin{rems}}
\newcommand{\erems}{\end{rems}}
\newtheorem{fact}{Fact}
\newcommand{\bfact}{\begin{fact}}
\newcommand{\efact}{\end{fact}}
\newtheorem{examp}{Example}
\newcommand{\bexamp}{\begin{examp}\rm}
\newcommand{\eexamp}{\end{examp}}
\newtheorem{defn}{Definition}
\newcommand{\bdefn}{\begin{defn}\rm}
\newcommand{\edefn}{\end{defn}}

\newtheorem{prob}{Problem}
\newcommand{\bprob}{\begin{prob}}
\newcommand{\eprob}{\end{prob}}

\newcommand{\bvtm}{\begin{verbatim}}
\newcommand{\bfig}{\begin{figure}}
\newcommand{\efig}{\end{figure}}
\newcommand{\bcen}{\begin{center}}
\newcommand{\ecen}{\end{center}}

\long\def\comment#1{}

\def \n2{{N_0 \over 2}}

\newcommand{\bP}[1]{{\mathbb{P}}\left[{#1}\right]}
\newcommand{\bE}[1]{{\mathbb{E}}\left[{#1}\right]}

\newcommand{\1}[1]{{\bf 1}\left[#1\right]}

\def \h5{\hspace{0.5in}}

\graphicspath{{Figures/}}

\begin{document}

\preprint{APS/123-QED}

\title{Robustness of power systems under a democratic fiber bundle-like model}

\author{Osman Ya\u{g}an}
%
\affiliation{%
 Department of ECE and CyLab, 
Carnegie Mellon University, Pittsburgh, PA 15213 USA}%


%

\date{\today}

\begin{abstract}
We consider a power system with $N$ transmission lines whose initial loads (i.e., power flows) $L_1, \ldots, L_N$ are 
independent and identically distributed with $P_L(x)=\bP{L \leq x}$. The {\em capacity}
 $C_i$ defines the maximum flow allowed on line $i$, and is assumed to be given by $C_i=(1+\alpha)L_i$, 
with $\alpha>0$. We study the robustness of this power system against {\em random} attacks (or, failures) that target a $p$-{\em fraction} of the lines,
under a {\em democratic} fiber bundle-like model. Namely, when a line fails, the load it was carrying is 
redistributed {\em equally} among the remaining lines. 
Our contributions are as follows: i) we show analytically that the final breakdown of the system always takes place through a first-order transition at 
the critical attack size $p^{\star}=1-\frac{\bE{L}}{\max_{x}(\bP{L>x}(\alpha x + \bE{L ~|~ L>x}))}$, where $\bE{\cdot}$ is the expectation operator; ii) we derive conditions on the distribution $P_L(x)$ for which the first order break down of the system occurs {\em abruptly} without any preceding {\em diverging} rate of failure; iii) we provide a detailed analysis of the robustness of the system under three specific load distributions: Uniform, Pareto,
and Weibull, showing that with the minimum load $L_{\textrm{min}}$ and mean load $\bE{L}$ fixed, Pareto distribution is the worst (in terms of robustness) among the three, whereas Weibull distribution is the best with shape parameter selected relatively large;
iv) we provide numerical results that confirm our mean-field analysis; and v) we show that $p^{\star}$ is maximized when the load distribution is a Dirac delta function centered at $\bE{L}$, i.e., when all lines carry the same load; we also show
that optimal
$p^{\star}$ equals $\frac{\alpha}{\alpha+1}$.  This last finding is particularly surprising given that
heterogeneity is known to lead to high robustness against {\em random} failures in many other systems.

\begin{description}
\item[PACS numbers]{64.60.Ht, 62.20.M-, 89.75.-k, 02.50.-r}


\end{description}
\end{abstract}
\maketitle

\section{Introduction}

As we embark on a future where the demand for electricity power is greater than ever, 
and the quality of life of the society highly depends on the  continuous functioning of 
power grid, a fundamental question arises as to how we can design a power system in a robust and reliable manner.
A major concern regarding such systems are the seemingly unexpected large scale failures. 
Although rare, the sheer size of such failures has proven to be very costly, 
at times affecting hundreds of millions of people \cite{rosas2011analysis,andersson2005causes}; e.g., the recent blackout in India \cite{zhang2013understanding,tang2012analysis}. 
Such events are often attributed to a 
small initial shock getting escalated due to intricate dependencies within a power system
\cite{Buldyrev,WattsExternal,kinney2005modeling}. 
This phenomenon, also known as cascade of failures, 
has the potential of collapsing an entire power system as well as
other infrastructures that depend on the power grid  \cite{o2007critical,dobson2007complex,YaganQianZhangCochranLong}; 
e.g., water, transport, communications, etc.
Therefore, understanding the dynamics of failures in power systems 
and mitigating the potential risks are critical for the successful
development and evolution of many critical infrastructures.


In this work, we study the robustness of power systems under a {\em democratic fiber bundle-like} model 
\cite{PahwaScoglioScala,Daniels1945,AndersenSornetteKwan}, which is based on the equal redistribution of load upon the failure of a power line. 
It was suggested by Pahwa et al. \cite{PahwaScoglioScala} that equal load redistribution can be a reasonable
assumption (in the mean-field sense) due to the long-range nature of Kirchoff's law.
This is especially so under  
the DC power flow model that approximates the standard AC power flow model 
when the phase differences along the branches are small and the bus voltages are 
fixed \cite{PahwaScoglioScala}. In many cases, power flow calculations based on the DC model is known 
\cite{Overbye,StottJardimAlsac} to give accurate results that match the AC model calculations. 


Our problem setting is as follows: We consider $N$ transmission lines whose initial loads (i.e., power flows) $L_1, \ldots, L_N$ are 
independently drawn from a distribution $P_L(x)=\bP{L \leq x}$. The maximum flow allowed on a line $i$ defines its {\em capacity}, and
is given by $C_i=(1+\alpha)L_i$ with $\alpha>0$ denoting the {\em tolerance} parameter. If a line fails (for any reason), its load will be redistributed
equally among all lines that are alive, meaning that the load carried by a line may
increase over time. We also assume that any line whose load exceeds its capacity will be tripped (i.e., disconnected) 
by means of automatic protective equipments so as to avoid costly damages to the system. 

We study the robustness of this system against {\em random} attacks (or, failures) that target a $p$-{\em fraction} of the lines. 
The failure of the $p$-fraction of lines may cause further failures in the system 
due to flows of some of the lines exceeding their capacity. Subsequently, their load will be redistributed which in turn may cause further failures, and so on until
the cascade of failures stops;  note that this process is guaranteed to converge, at the very least when {\em all} lines in the system fail.

One of our
important findings is to show the existence of
a critical threshold on the attack size $p$, denoted by $p^{\star}$, below which a
considerable fraction lines remain functional
at the steady state; on the other hand, if $p>p^{\star}$, the entire system collapses.
We show that the critical attack size is given by $p^{\star}=1-\frac{\bE{L}}{\max_{x}(\bP{L>x}(\alpha x + \bE{L ~|~ L>x}))}$, where $\bE{\cdot}$ denotes the expectation operator.
In addition, we show that 
the {\em phase transition} at $p^{\star}$ is always {\em first-order}; i.e., the variation
of the \lq \lq fraction of functional lines at the steady state" with respect to \lq\lq attack size $p$" has a discontinuous first derivative.
In a nutshell, what this means is that power systems under the democratic fiber bundle model 
tend to exhibit very large changes to small variations on the failure size (around $p^{\star}$), rendering their robustness unpredictable from previous data. 
In fact, this type of first order phase transition is attributed \cite{WattsExternal} to be the origin of large but rare blackouts seen in real world, in a way explaining
how small initial shocks can cascade to collapse
large systems that have proven stable with respect to similar  
disturbances in the past.

Our second main contribution is to demonstrate the clear distinction 
between the case where the first order break down of the system occurs {\em abruptly} without any preceding {\em diverging} rate of failure versus the case where
a second order transition precedes the first-order breakdown. In the former case, if  $p<p^{\star}$
the final fraction of alive lines will be given by $1-p$ meaning that no single additional line fails other than those that are initially attacked, whereas the whole system 
will suddenly 
collapse if the attack size exceeds $p^{\star}$. 
These cases are reminiscent of the most catastrophic and unexpected large-scale collapses
observed in the real world. We provide explicit conditions on the distribution $P_L(x)$ of the loads 
and the tolerance parameter $\alpha$ that distinguish the two cases. 

Last but not least, we show that $p^{\star}$ is maximized when the load distribution is a Dirac delta function centered at $\bE{L}$, i.e., when all lines carry the same load. 
The optimal $p^{\star}$ is shown to be given by $\frac{\alpha}{\alpha+1}$, regardless of the mean load $\bE{L}$. 
This finding is particularly surprising given that
complex networks are known to be extremely robust against random failures when their degree distribution is broad \cite{BarabasiAlbert};
 e.g., when the number of links incident on a line follows a power-law distribution.

We believe that our results provide interesting insights into the dynamics of 
cascading failures in power systems. In particular, they can help design power systems in a more robust manner. 
The results obtained here may have applications in fields other than power systems as well. Fiber bundle models have been used in a wide range of applications including 
fatigue \cite{curtin1993tough}, failure of composite materials \cite{kun2007fatigue}, 
landslides \cite{cohen2009fiber}, etc. A particularly interesting application is the study of the traffic jams 
in roads \cite{pradhan2003failure}, where the capacity of a line can be regarded as the traffic flow capacity of a road.

The paper is structured as follows. In Section \ref{sec:Model} we give the 
details of our system model, discuss how it compares with other models in the literature, and comment
on its applicability in power systems. Analytical results 
regarding the 
robustness of the system against random attacks are provided in Section \ref{sec:Results} for general load distributions.
These results are discussed in more details for  three specific load distributions in Section \ref{sec:distributions} and various load-distribution-specific conclusions are drawn.
Section \ref{sec:Simu} is devoted to numerical results 
that confirm the main findings of the paper for systems of finite size. In Section \ref{sec:optimal}, we derive the {\em optimal} load distribution 
that leads to maximum robustness among all distributions with the same mean, and the paper is concluded 
in Section \ref{sec:Conclusion}.

\section{Model definitions}
\label{sec:Model}
We consider a power system with $N$ transmission lines whose initial loads (i.e., power flows) $L_1, \ldots, L_N$ are 
independent and identically distributed with $P_L(x):=\bP{L \leq x}$. The corresponding probability density
function is given by $p_L(x) = \frac{d}{dx}P_L(x)$.
Let $L_{\textrm{min}}$ denote the minimum value $L$ can take; i.e.,
\[
L_{\textrm{min}} = \sup\{x: P_L(x) = 0\}.
\]
We assume that $L_{\textrm{min}} > 0$. We also assume that the density $p_L(x)$ is continuous on its support.

The {\em capacity} of a line defines the maximum power flow that it can sustain, and 
is typically \cite{MotterLai,WangChen,Mirzasoleiman,CrucittiLatora} set to be a fixed factor of the line's original load.
To that end, we let the capacity $C_i$ of line $i$ be given by  
\begin{equation}
C_i=(1+\alpha)L_i, \qquad i=1,\ldots, N,
\label{eq:capacity}
\end{equation}
with $\alpha>0$ defining the {\em tolerance} parameter. For simplicity, we assume that all lines have the same tolerance parameter $\alpha$, 
but it would be of interest to extend our results to the case
where the tolerance parameter $\alpha_i$ of a line $i$ is randomly selected from a  probability distribution, for each $i=1,\ldots,N$.
A line {\em fails} (i.e., outages) if its load exceeds its capacity at any given time. In that case, the load it was carrying before the failure
is redistributed {\em equally} among all remaining lines. 

Our main goal is to study the robustness of this power system against 
{\em random} attacks that result with a failure of a $p$-fraction of the lines; of course, all the discussion and 
accompanying results do hold for the robustness against random {\em failures} as well.
The initial set of failures leads to redistribution of power flows from the failed lines to {\em alive} ones (i.e., non-failed lines), 
so that the load on each alive line becomes equal to its initial load plus its equal share of the total load of the failed lines.  This may lead to 
the failure of some additional lines due to 
the updated flow exceeding their capacity. This process may continue recursively, generating a {\em cascade of failures}, 
with each failure further increasing the load on the alive lines, and may eventually result
with the collapse of the entire system. Throughout, we let $n_{\infty}(p)$ denote the {\em final} fraction of alive lines
when a $p$-fraction of lines is randomly attacked. The robustness of a power system will be evaluated 
by the behavior of $n_{\infty}(p)$ as the attack size $p$ increases, and particularly by the critical attack size $p^{\star}$ at which 
$n_{\infty}(p)$ drops to zero.

Our formulation is partially inspired by the democratic fiber bundle model \cite{AndersenSornetteKwan,Daniels1945},
where $N$ parallel fibers with random failure thresholds $C_1, \ldots, C_N$ (i.e., capacities) drawn independently from $P_C(x)$ share equally an applied total force of $F$; see also 
\cite{Silveira,sornette1997conditions,pradhan2003failure,roy2015fiber}. This model has been recently adopted by Pahwa et al. \cite{PahwaScoglioScala} in the context 
of power systems
with $F$ corresponding to the total load that $N$ power lines share equally.
A major difference of our setting with the original democratic fiber-bundle model is that in the latter the total load of 
 the system is always fixed at $F$. This ensures that the load that each alive line carries at any given time is {\em independent}
of the specific set of lines that have failed until that time. 
For example if $M$ lines out of the original $N$ are alive, 
one can easily compute the load per alive line as $F/M$ regardless of which 
$N-M$ lines have actually failed. In our model, however,
the initial loads of $N$ lines are random and they differ from each other, and so do their capacities.
This leads to strong dependencies between the load of an alive line and the particular
$N-M$ set of lines that have failed, and makes it impossible to compute 
the former merely from the
number of failed lines.
For instance, at any given time, lines that are alive are likely to have a larger capacity, and thus a larger initial load in 
view of (\ref{eq:capacity}), than those that have failed. In addition, the 
 total load shed on to the alive lines is not given by $(N-M)\bE{L}$, since 
the lines that have failed are likely to have a smaller capacity, and thus a smaller initial load, than average. 
As a result of these intricate dependencies, analysis of  cascading failures in our setting becomes substantially 
more challenging than that in the fiber-bundle model; see Section \ref{sec:Results} for details.

We believe that our problem formulation can lead to significant insights for the robustness of power systems (and possibly of other real-world systems) 
that can not be seen in the original fiber-bundle model. First of all, our formulation allows analyzing the robustness of the system against external attacks
or random line failures, which are known to be the source of system-wide blackouts in many interdependent systems \cite{Rosato,Buldyrev,YaganQianZhangCochranLong}; the 
standard fiber-bundle model
is instead concerned with failures triggered by increasing the total force (i.e., load) applied to the system.
Secondly, unlike the democratic fiber bundle model where all lines start with the same initial load \footnote{The case of non-uniform initial loads in democratic fiber bundle model is briefly discussed 
in \cite{PahwaScoglioScala} in the context of power systems, and some numerical results are provided}, power lines in real systems are likely to have different loads at the initial set-up although 
they may participate equally in taking over the load of those lines that have failed;
intuitively speaking, this is also the case for traffic flow on roads.

Our model has some similarities also with the CASCADE model introduced by Dobson et al. \cite{dobson2002examining}.
There, they assume that initial loads $L_1,\ldots,L_N$ are uniformly distributed over an interval $(L_{\textrm{min}},L_{\textrm{max}})$,
and all lines have the same capacity $C =L_{\textrm{max}}$. This is a significant difference from our model where capacities vary according to (\ref{eq:capacity}).
Another major difference is that in the CASCADE model, a fixed amount $\Delta$ is redistributed to all alive lines irrespective of the load being carried before failure. Therefore, strong
dependencies between particular lines failed and the load carried by alive lines do not exist in the CASCADE model.

A word on notation in use: The random variables (rvs)
under consideration are all defined on the same probability space
$(\Omega, {\cal F}, \mathbb{P})$. Probabilistic statements are
made with respect to this probability measure $\mathbb{P}$, and we
denote the corresponding expectation operator by $\mathbb{E}$.
The indicator function of an event $A$ is denoted by $\1{A}$. 
  
\section{Analytic Results}
\label{sec:Results}

\subsection{Recursive Relations}
We now provide the mean-field analysis of the cascading failures of lines for the model described in Section \ref{sec:Model}.
We start by deriving recursive relations concerning the fraction $f_t$ of lines that are {\em failed} at time stage $t=0,1,\ldots$. 
The number of links that are still alive at time $t$ is then given by $N_t = N (1-f_t)$ for all $t=0,1,\ldots$. The cascading failures start with 
a random attack that targets a fraction $p$ of power lines, whence we have $f_0 = p$. Upon the failure of these $f_0 p$ lines, their load will be redistributed
to the remaining $(1-f_0)N$ lines. The resulting extra load per alive line, $Q_0$ is given by 
\begin{equation}
Q_0 =  \frac{\bE{L} p N }{(1-p) N} = \bE{L} \frac{f_0}{1-f_0}.
\label{eq:Q_0}
\end{equation}
At this initial stage, since the $pN$ lines that have been attacked are selected uniformly at random, the mean total load that will be transferred to the remaining lines is just given by
$\bE{L} p N$. 

Now, in the next stage a line $i$ that survived the initial attack will fail if and only if its new load reaches its capacity \footnote{For convenience, we assume that a line also fails when its load
equals its capacity.}; i.e., if 
\[
 L_i + Q_0 \geq (1+\alpha)L_i, 
\]
or, equivalently if $L_i \leq Q_0/\alpha$. Therefore, at stage $t=1$, an additional fraction $\bP{L \leq Q_0/\alpha}$ of lines will fail from the lines that were alive at the end of stage 0.
This gives
\[
f_1= f_0 + (1-f_0) \bP{L \leq Q_0/\alpha} = 1-(1-f_0) \bP{L > \frac{Q_0}{\alpha}}.
\]

In order to compute $Q_1$, i.e., the total extra load per alive line at stage $1$, we should sum 
the total load of all the failed lines until this stage and divide it by the new system size $1-f_1$. So, $Q_1 (1-f_1)$ is given by the sum of $Q_0 (1-f_0)$,
and the total load of the lines failed {\em at stage} 1 normalized by the number
of lines $N$; i.e., of lines that survived the initial attack but have load $L \leq Q_0/\alpha$. Let $\mathcal{A}$ be the initial set of lines
attacked. We get
\begin{eqnarray}
Q_1 (1-f_1) &=& Q_0 (1-f_0)+ \frac{1}{N} \cdot \bE{\sum_{i \not \in \mathcal{A}: L_i \leq Q_0/\alpha} L_i}
\nonumber \\
&=& Q_0 (1-f_0) + \frac{1}{N} \cdot \bE{\sum_{i\not \in \mathcal{A}} L_i \1{L_i \leq Q_0/\alpha}}
\nonumber \\
&=& Q_0 (1-f_0) + \frac{1}{N} \cdot  \sum_{i\not \in \mathcal{A}} \bE{L_i \1{L_i \leq Q_0/\alpha}}
\nonumber \\
&=& Q_0 (1-f_0) + (1-f_0) \bE{L \1{L \leq Q_0/\alpha} },
\nonumber
\end{eqnarray}
where the last step uses $|\mathcal{A}|/N=p=f_0$. Thus, we get
\[
Q_1 =\frac{p \bE{L} + (1-p)\bE{L \cdot \1{L \leq Q_0/\alpha}}}{1-f_1}
\]
upon noting (\ref{eq:Q_0}). We find it useful to note that 
\[
\bE{L \cdot \1{L \leq Q_0/\alpha}} = \bE{L ~|~ L \leq Q_0/\alpha} \bP{L \leq Q_0/\alpha}.
\]

The general form of $f_t$ and $Q_t$ will become apparent as we compute them at stage
$t=2$. This time we argue as follows. For a line to still stay alive at this stage, two conditions need to be satisfied: i) it should not have failed
until this stage, which happens with probability $1-f_1$; and ii) its load should satisfy $L > Q_1/\alpha$ so that its capacity is still larger than its current load. One additional 
note is that a line that satisfies condition (i) necessarily have a load $L > Q_0/\alpha$. Collecting, we obtain
\[
f_2 = 1 - (1-f_1) \bP{L > Q_1/\alpha ~|~ L > Q_0/\alpha}.
\]
The total load that will be redistributed to the remaining lines can then be computed as before:
\begin{eqnarray}
Q_2 (1-f_2) &=& Q_0 (1-f_0)+ \frac{1}{N} \cdot \bE{\sum_{i \not \in \mathcal{A}: L_i \leq Q_1/\alpha} L_i}
\nonumber \\
&=& Q_0 (1-f_0) + (1-f_0) \bE{L \1{L \leq Q_1/\alpha} }.
\nonumber
\end{eqnarray}
One can complicate the matters a little bit and get the same expression by writing
\[
Q_2 (1-f_2) = Q_1 (1-f_1)+ \frac{1}{N} \bE{\sum_{i \not \in \mathcal{A}: Q_0/\alpha < L_i \leq Q_1/\alpha} L_i}
\]
as well. 

The form of the recursive equations is now clear. Let $f_0 = p$, $N_0 = N (1-p)$, and $Q_0= \bE{L} \frac{p}{1-p}$. For convenience, also
let $Q_{-1} = 0$. Then, for each $t=0,1, \ldots$, we have
\begin{equation}
\begin{array}{ll}
 f_{t+1} = &1- (1-f_t) \bP{L > \frac{Q_t}{\alpha} ~\bigg|~ L > \frac{Q_{t-1}}{\alpha}}\\
 Q_{t+1} =&\frac{p \bE{L} + (1-p) \bE{L \cdot \1{L \leq  \frac{Q_t}{\alpha}}}}{1-f_{t+1}}\\
 & \\
 N_{t+1} =&(1-f_{t+1}) N      
\end{array}
\label{eq:recursion}
\end{equation}

From (\ref{eq:recursion}) we see that cascades stop and a steady is reached, i.e., $N_{t+2}=N_{t+1}$, if 
\begin{equation}
\bP{L > \frac{Q_{t+1}}{\alpha} ~\bigg|~ L > \frac{Q_{t}}{\alpha}} =1.
\label{eq:cond_steady_state}
\end{equation} 
In order to understand the conditions that would lead to 
(\ref{eq:cond_steady_state}), we need to simplify the recursion on $f_t$. This step is taken in the next section.

\subsection{Conditions for steady-state via a simplification}
Applying the first relation in (\ref{eq:recursion}) repeatedly, we see that
\begin{eqnarray}\nonumber
\begin{array}{ll}
1-f_{t+1} &= (1-f_t) \bP{L > {Q_t}/{\alpha} ~|~ L > {Q_{t-1}}/{\alpha}} \\
1-f_{t} &= (1-f_{t-1}) \bP{L > {Q_{t-1}}/{\alpha} ~|~ L > {Q_{t-2}}/{\alpha}} \\
~~~\vdots & \\
1-f_{1} &= (1-f_0) \bP{L > Q_0/\alpha} 
\end{array}
\end{eqnarray}
Applying these recursively, we obtain
\[
1-f_{t+1}  = (1-f_0) \prod_{\ell = 0}^{t} \bP{L > {Q_{\ell}}/{\alpha} ~|~ L > {Q_{\ell-1}}/{\alpha}},
\]
where $Q_{-1} = 0$ as before. Since $Q_t$ is monotone increasing in $t$, i.e., $Q_{t+1}\geq Q_t$ for all $t$, we further obtain
\begin{align}
\lefteqn{1-f_{t+1}} &
\nonumber \\
 & =  (1-f_0) \frac{\bP{L > \frac{Q_{t}}{\alpha}}}{\bP{L > \frac{Q_{t-1}}{\alpha}}} \cdot \frac{\bP{L > \frac{Q_{t-1}}{\alpha}}}{\bP{L > \frac{Q_{t-2}}{\alpha}}} \cdots \frac{\bP{L > \frac{Q_{1}}{\alpha}}}{\bP{L > \frac{Q_{0}}{\alpha}}} 
 \nonumber \\
  & ~~~~~~~ \cdot \bP{L > \frac{Q_{0}}{\alpha}}
   \nonumber \\
  & = (1-f_0)\bP{L > {Q_{t}}/{\alpha}} 
  \label{eq:simplified_f_t}
\end{align}

Reporting this into (\ref{eq:recursion}) and recalling that $f_0=p$, we get the following simplified recursions:
\begin{equation}
\begin{array}{ll}
 f_{t+1} = &1- (1-f_t)  \bP{L > \frac{Q_t}{\alpha} ~|~ L > \frac{Q_{t-1}}{\alpha}}\\
 Q_{t+1} =& \frac{p \bE{L} + (1-p) \bE{L \cdot \1{L \leq  \frac{Q_t}{\alpha}}}}{(1-p) \bP{L > {Q_{t}}/{\alpha}}}\\
 & \\
 N_{t+1} =&(1-p) \bP{L > {Q_{t}}/{\alpha}} N
\end{array}
\label{eq:recursion_2}
\end{equation}
Failures will stop and a steady-state will be reached when $f_{t+2}=f_{t+1}$. From the first
relation in (\ref{eq:recursion_2}), we see that this holds if 
\[
\bP{L > {Q_{t+1}}/{\alpha} ~|~ L > {Q_{t}}/{\alpha}} =1,
\]
or, equivalently if
\begin{align}
 &  \bP{L >  \frac{p \bE{L}  + (1-p) \bE{L \cdot \1{L \leq  \frac{Q_t}{\alpha}}}}{\alpha(1-p) \bP{L > {Q_{t}}/{\alpha}}} ~\Bigg|~ L > \frac{Q_{t}}{\alpha}} 
\nonumber 
 \\
 & ~= 1,
 \label{eq:stop_condition} 
\end{align}
as we use the middle equation in (\ref{eq:recursion_2}).

Define $x:=Q_t/\alpha$, and realize that
\begin{align}\nonumber
& p \bE{L}  + (1-p) \bE{L \cdot \1{L \leq  x}}  \\ \nonumber
 & =p \bE{L}  + (1-p) \bE{L \cdot (1-\1{L >  x})}  \\  \nonumber
 & = \bE{L} - (1-p) \bE{L \cdot \1{L >  x}}.
\end{align}
With these in place, the condition for cascades to stop (\ref{eq:stop_condition}) gives 
 \begin{align}
 \bP{L >  \frac{\bE{L} - (1-p) \bE{L \cdot \1{L >  x}}}{\alpha(1-p) \bP{L >x}} ~\Bigg|~ L > x} =1.
 \label{eq:stop_condition_2} 
 \end{align}
It is now clear how to obtain the final fraction of power lines that are still alive at the end of the cascading failures: 
One must find the smallest  solution $x^{\star}$ of (\ref{eq:stop_condition_2}). Then, the final fraction $n_{\infty}(p)$ of alive lines is given (see (\ref{eq:simplified_f_t})) by
\begin{equation}
n_{\infty}(p) = 1 - f_{\infty} = (1-p) \bP{L > x^{\star}}.
\label{eq:final_size}
\end{equation}

Under the enforced assumptions on the distribution of $L$, we see that (\ref{eq:stop_condition_2}) holds in either one of the following cases:
\begin{itemize}
\item[$i)$] If $x \geq \frac{\bE{L} - (1-p) \bE{L \cdot \1{L >  x}}}{\alpha(1-p) \bP{L >x}}$; or,
\item[$ii)$] If $x < \frac{\bE{L} - (1-p) \bE{L \cdot \1{L >  x}}}{\alpha(1-p) \bP{L >x}}$ and 
\begin{equation}
\bP{L >  \frac{\bE{L} - (1-p) \bE{L \cdot \1{L >  x}}}{\alpha(1-p) \bP{L >x}}}=1.
\label{eq:second_option}
\end{equation}
\end{itemize}
We see that in the latter case, it automatically holds $\bP{L >  x}=1$, meaning that the final system size equals
$1-p$. In other words, no single line fails other than the $pN$ lines that went down as a result of the initial attack. 
Using $\bP{L >  x}=1$ in (\ref{eq:second_option}), we see that this happens whenever 
$\bP{L >  \frac{p\bE{L}}{\alpha(1-p)}} = 1$, which can be regarded as the {\em condition for no cascade of failures}.
This condition can help in capacity provisioning, i.e., in determining the factor $\alpha$ needed for robustness against $p$-size attacks, and can be rewritten 
as 
\begin{equation}
L_{\textrm{min}} >  \frac{p\bE{L}}{\alpha(1-p)}.
\label{eq:cond_on_L_min}
\end{equation}

The first condition, on the other hand, amounts to 
\begin{equation}
\bP{L > x} \left(\alpha x + \bE{L ~|~ L > x} \right) \geq \frac{\bE{L}}{1-p}.
\label{eq:main_condition}
\end{equation}
We can now see that the final system size $n_{\infty}(p)$ is {\em always} given by $(1-p) \bP{L > x^{\star}}$ where $x^{\star}$ is the smallest solution of (\ref{eq:main_condition}).
This is clearly true for the case $(i)$ given above.
To see why this approach also works for the case $(ii)$, observe that when (\ref{eq:cond_on_L_min}) 
holds (\ref{eq:main_condition}) is satisfied for any $x$ 
in $[\frac{p\bE{L}}{\alpha(1-p)} , L_{\textrm{min}} ]$. Hence, the smallest solution $x^{\star}$
of (\ref{eq:main_condition}) will always give $x^{\star} \leq L_{\textrm{min}}$, leading to
$(1-p) \bP{L > x^{\star}} = 1-p$. As discussed before, no cascade takes place
under (\ref{eq:cond_on_L_min}) (i.e., in the case $(ii)$ above), so the final system size
is indeed $1-p$.

For a graphical solution of $n_{\infty}(p)$, 
one shall plot $\bP{L > x} \left(\alpha x + \bE{L ~|~ L > x} \right)$ as a function of $x$ 
(e.g., see Figure \ref{fig:gamma_example}), and draw a horizontal line at the height $\bE{L}/(1-p)$ on the same plot. The leftmost intersection of these two lines gives the operating point $x^{\star}$, from which we can compute  $n_{\infty}(p) = (1-p)\bP{L > x^{\star}}$. When there is no intersection,  we set $x^{\star}=\infty$ and understand that $n_{\infty}(p)=0$.

\begin{figure}[!h]
\centering\subfigure[]{\hspace{-0.5cm} \includegraphics[totalheight=0.3\textheight,
width=.5\textwidth] {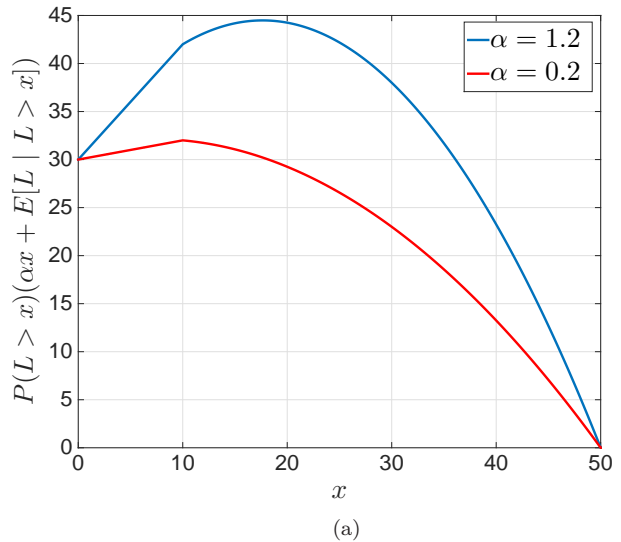} \label{fig:gamma_example}}
\subfigure[]{\hspace{-0.5cm}
\includegraphics[totalheight=0.3\textheight,
width=.5\textwidth] {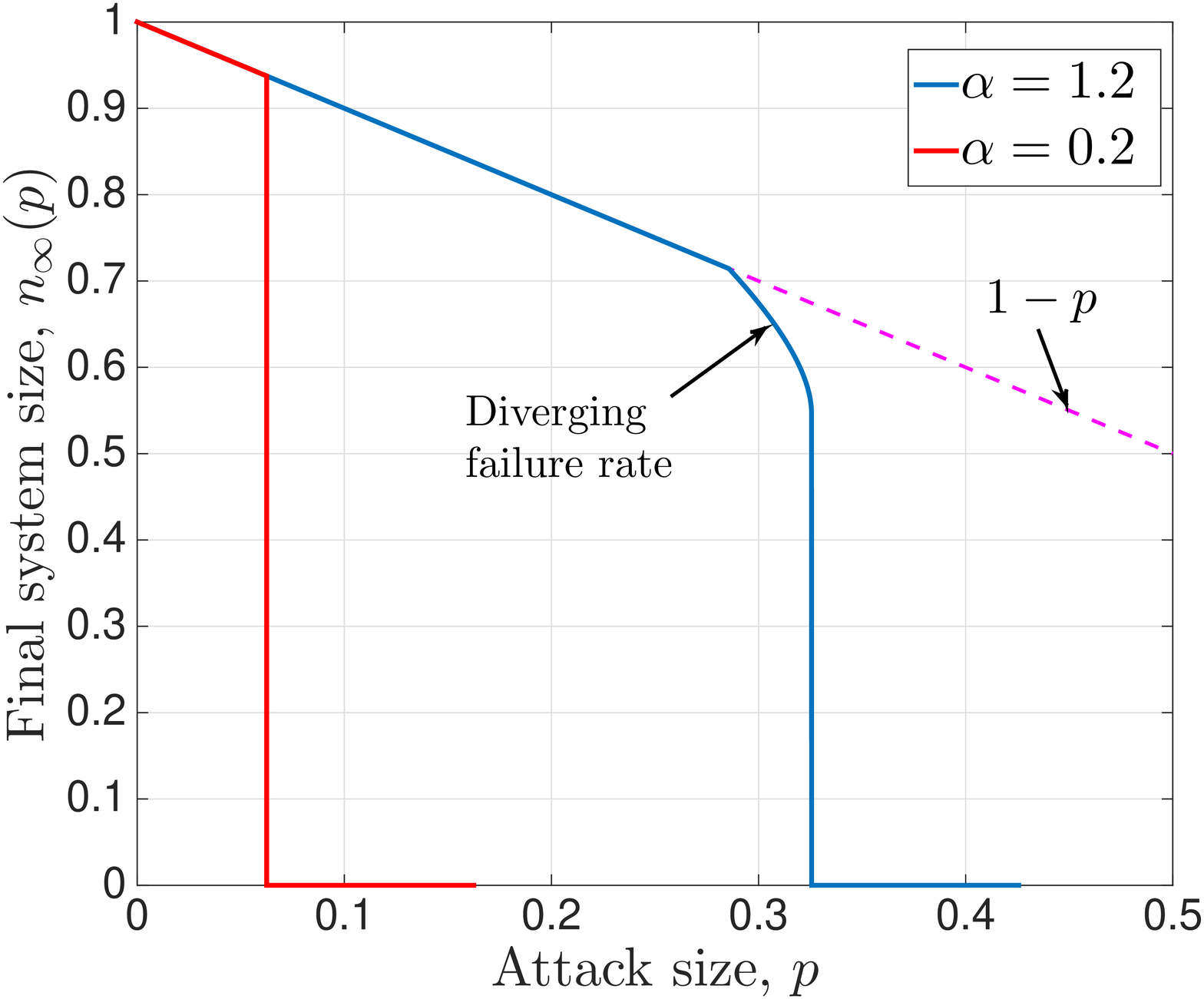}\label{fig:n_inf_example}}
 \caption{ \sl (Color online) We demonstrate the distinction between an abrupt first-order rupture, and a first-order rupture
 that is preceded by a diverging failure rate. $p_L(x)$ is assumed to be of uniform density over the range
 $[L_{\textrm{min}},L_{\textrm{max}}] = [10,50]$. In both plots, Red curves stand for the case where $\alpha = 0.2$, whereas 
 Blue curves represent $\alpha=1.2$. Figure \ref{fig:gamma_example} shows $\bP{L > x} \left(\alpha x + \bE{L ~|~ L > x} \right)$, whereas 
 Figure \ref{fig:n_inf_example} plots the corresponding variation of $n_{\infty}(p)$ with attack size $p$. We observe that for $\alpha=0.2$ (Red),
 $\bP{L > x} \left(\alpha x + \bE{L ~|~ L > x} \right)$ takes its maximum at the point $x=L_{\textrm{min}} = 10$. As a result, 
 we see an abrupt first-order transition of $n_{\infty}(p)$ as it suddenly drops to zero at the point $p=p^{\star} = 0.0625$, while decaying linearly as $1-p$ up until that point.
 The case where $\alpha=1.2$ is clearly different as  $\bP{L > x} \left(\alpha x + \bE{L ~|~ L > x} \right)$ is now maximized at $x=17.6 > L_{\textrm{min}} $. As expected from our discussion,
 this ensures that the total failure of the system occurs {\em after} a diverging failure rate is observed. This divergence is clearly seen in Figure \ref{fig:n_inf_example}  where the dashed line
 corresponds to the $1-p$ curve.
  }
\label{fig:examples}
\end{figure}

\subsection{Rupture Condition}
We now know how to compute the final system size $n_{\infty} (p)$ for a given attack size $p$. In many cases, we will be interested in the 
variation of $n_{\infty} (p)$ as a function of $p$. This will help us understand the response of the system to attacks of varying magnitude. 
Of particular interest will be to derive the critical attack size $p^{\star}$ such that for any attack with size $p > p^{\star}$, the system undergoes a complete breakdown leading to 
$n_{\infty} (p) = 0$. 

From (\ref{eq:main_condition}) and the discussion that follows, we see that the maximum attack size $p^{\star}$ is related to the {\em global} maximum of the function
$\bP{L > x} \left(\alpha x + \bE{L ~|~ L > x} \right)$. In fact, it is easy to see that
\begin{equation}
p^{\star} = 1- \frac{\bE{L}}{\max\limits_{x}\{\bP{L > x} \left(\alpha x + \bE{L ~|~ L > x} \right)\}}.
\label{eq:max_attack}
\end{equation} 
The critical point $x^{\star}$ that maximizes the function $\bP{L > x} \left(\alpha x + \bE{L ~|~ L > x} \right)$ can shed light on the type 
of the transition that the system undergoes as the attack size increases. First of all, the system will always undergo a 
{\em first-order} (i.e., discontinuous) transition at the point $p^{\star}$. This can be seen as follows: We have $n_{\infty} (p^{\star^{+}}) = 0$ by virtue of the fact that
no $x$ will satisfy (\ref{eq:main_condition}), and cascading failures will continue until the whole system breaks down. 
On the other hand, $n_{\infty} (p^{\star^{-}}) = (1-p) \bP{L > x^{\star}}>0$ 
where $x^{\star}$ is the point that maximizes $\bP{L > x} \left(\alpha x + \bE{L ~|~ L > x} \right)$. 
We can see why it must hold $\bP{L > x^{\star}}>0$ via contradiction:
 $\bP{L > x^{\star}}=0$ implies that the maximum value of $\bP{L > x} \left(\alpha x + \bE{L ~|~ L > x} \right)$ is zero, which clearly does not hold since at $x=0$ this function equals $\bE{L}>0$ by non-negativity of $L$.

An interesting question is whether this first order rupture at the point $p^{\star}$ will have any early indicators at smaller attack sizes; e.g., a 
{\em diverging} failure rate leading to a non-linear decrease in  $n_{\infty} (p)$. With $L_{\textrm{min}} > 0$, we know from (\ref{eq:cond_on_L_min}) that for $p$ sufficiently small, 
there will be no cascades and $n_{\infty} (p)$ will decrease linearly as $1-p$. 
This corresponds to the situations where (\ref{eq:main_condition}) is satisfied at a point $x \leq L_{\textrm{min}}$, i.e., when 
 $\bP{L > x} \left(\alpha x + \bE{L ~|~ L > x} \right)$ is linearly increasing with $x$.
An {\em abrupt} first-order transition is said to take place if the linear decay of $n_{\infty} (p)$ is followed by a sudden discontinuous jump to zero at the point $p^{\star}$.  
Those cases are reminiscent of the real-world phenomena 
of unexpected large-scale system collapses; i.e., cases where seemingly identical attacks/failures leading to entirely different consequences. 

It is easy to see that an abrupt transition occurs if $\bP{L > x} \left(\alpha x + \bE{L ~|~ L > x} \right)$ takes its maximum at the point $x=L_{\textrm{min}}$; see Figure \ref{fig:examples}. In that case, (\ref{eq:main_condition}) either has a solution at some $x \leq L_{\textrm{min}}$ 
so that $n_{\infty} (p)=1-p$, or has no solution leading to $n_{\infty} (p)=0$. 
Under the assumptions enforced here, $\bP{L > x} \left(\alpha x + \bE{L ~|~ L > x} \right)$  is continuous at every $x\geq0$. Given that this function is linear increasing on the range $0 \leq x \leq L_{\textrm{min}}$, a maximum
takes place at $x=L_{\textrm{min}}$ if at that point the derivate changes its sign. 
We have
\begin{align}
&\frac{d}{dx} \left(\bP{L > x} \left(\alpha x + \bE{L ~|~ L > x} \right) \right)  
\\ \nonumber
& = \frac{d}{dx} \left( \alpha x \bP{L > x} +   \bE{L \cdot \1{L > x}}  \right)  
\\ \nonumber
& = \alpha \bP{L > x}  + \alpha x (-p_L(x)) + \frac{d}{dx} \left(\int_{x}^{\infty} t p_L(t)dt \right) 
\\ \nonumber
& =  \alpha \bP{L > x}  + \alpha x (-p_L(x)) - x p_L(x)
\\ \label{eq:condition_derivative}
& = \alpha \bP{L > x} - x p_L(x) (\alpha + 1)
\end{align}
where in the second to last step we used the Leibniz integral rule. As expected, for $x < L_{\textrm{min}}$, we have 
$\bP{L > x}=1$ and $p_L(x)=0$, so that the derivative is constant at $\alpha$. For an abrupt rupture to take place, 
the derivative should be negative at the point $x=L_{\textrm{min}}$; i.e., we need 
\[
\alpha -  L_{\textrm{min}} \cdot p_L( L_{\textrm{min}}) (\alpha +1) < 0,
\] 
or, equivalently
\begin{equation}
\frac{\alpha}{(\alpha+1) L_{\textrm{min}}} < p_L( L_{\textrm{min}}).
\label{eq:rupture_condition}
\end{equation}

It is important to note that (\ref{eq:rupture_condition}) ensures only the existence of a local maximum 
of the function $\bP{L > x} \left(\alpha x + \bE{L ~|~ L > x} \right)$ at the point $x=L_{\textrm{min}}$. This in turn 
implies that there will be a first order jump in $n_{\infty} (p)$ at the point
where $\bE{L}/(1-p) =  \alpha L_{\textrm{min}} + \bE{L} $; i.e., at the point $p$ that satisfies (\ref{eq:cond_on_L_min}) with 
equality. However, for this condition to lead to an \lq\lq abrupt" first-order {\em breakdown}, we need $x=L_{\textrm{min}}$ to be the {\em global} maximum.
This can be checked by finding all $x$ that make the derivate at (\ref{eq:condition_derivative}) zero, and then comparing the corresponding maximum points. 
If $x=L_{\textrm{min}}$ is only a local maximum, then the system will have a sudden drop in size at the corresponding attack size, but will not undergo a complete 
failure; the complete failure and the drop of $n_{\infty} (p)$ to zero will take place at a larger attack size where, again there will be a first-order transition; e.g., see Figure \ref{fig:2_step}.

We close by giving the general condition for first-order jumps to take place. We need a change of sign of the derivative at (\ref{eq:condition_derivative}), leading to 
\[
\pm \alpha \bP{L > x} - x p_L(x) (\alpha + 1) \bigg|_{x=x^{\star^{\pm}}} < 0.
\]
Equivalently, a first-order jump will be seen for every $x^{\star}$ satisfying
\begin{equation}
p_L(x^{\star^{-}}) < \frac{\alpha \bP{L > x^{\star}}}{(\alpha+1) x^{\star}} < p_L(x^{\star^{+}}).
\end{equation}

\section{Results with specific distributions} 
\label{sec:distributions}
We analyze a few specific distributions in more details. Namely, we will consider 
Uniform, Pareto, and Weibull distributions. 

\subsection{Uniform distribution} 
Assume that loads $L_1,\ldots,L_N$ are uniformly distributed over $[ L_{\textrm{min}},  L_{\textrm{max}}]$. 
In other words, we have 
\[
p_L(x)=\frac{1}{ L_{\textrm{max}}  - L_{\textrm{min}} } \cdot \1{ L_{\textrm{min}} \leq x  \leq L_{\textrm{max}} },
\]
so that
\begin{align}
\bP{L > x} &= \frac{ L_{\textrm{max}} -x }{ L_{\textrm{max}} -  L_{\textrm{min}} } \1{ L_{\textrm{min}} \leq x  \leq L_{\textrm{max}}} 
\nonumber \\
&~~~~+ \1{ x< L_{\textrm{min}} }.
\end{align}
We see that over the range $x$ in $[0, L_{\textrm{max}})$,
the derivative of  $\bP{L > x} \left(\alpha x + \bE{L ~|~ L > x} \right)$ 
(see (\ref{eq:condition_derivative})) is either never zero or becomes zero only once 
at  
\[
x^{\star} = \frac{\alpha}{2\alpha+1}L_{\textrm{max}},
\]
For the latter to be possible, we need $\frac{\alpha}{2\alpha+1}L_{\textrm{max}} \geq L_{\textrm{min}}$. 
If the opposite condition holds, i.e., if $\frac{\alpha}{2\alpha+1}L_{\textrm{max}} < L_{\textrm{min}}$, then 
$\bP{L > x} \left(\alpha x + \bE{L ~|~ L > x} \right)$  is maximized at $x=L_{\textrm{min}}$, and an {\em abrupt} first order break down 
will occur (as $p$ increases) without any preceding diverging failure rate. As expected, the condition $\frac{\alpha}{2\alpha+1}L_{\textrm{max}} < L_{\textrm{min}}$ is equivalent
to the general rupture condition (\ref{eq:rupture_condition}) and can be written most compactly as
\[
\alpha < \frac{L_{\textrm{min}}}{\max\left(L_{\textrm{max}}-2L_{\textrm{min}},~ 0 \right)}.
\]
It  follows that if $L_{\textrm{max}} \leq 2 L_{\textrm{min}}$, then an abrupt rupture takes place 
irrespective of the tolerance factor $\alpha$.

\subsection{Pareto distribution} 
Distribution of many real world variables are shown to exhibit a {\em power-law} behavior,
with very large variability  \cite{faloutsos1999power,ClausetShaliziNewman,newman2002spread,LeichtDSouza}. 
To consider power systems where the initial loads of the lines can exhibit high variance, we consider the case where $L_1, \ldots, L_N$ are drawn from a Pareto distribution: Namely, with
$b,L_{\textrm{min}}>0$, we set
\[
p_L(x) =  L_{\textrm{min}}^{b} b x^{-b-1} \1{x \geq L_{\textrm{min}}}.
\]
To ensure that $\bE{L}$ is finite, we also enforce that $b>1$; in that case we have $\bE{L}=\frac{b  L_{\textrm{min}}}{b-1}$. Then, the condition for an abrupt first order rupture (\ref{eq:rupture_condition}) gives
\[
\frac{\alpha}{(\alpha+1) L_{\textrm{min}}} <   L_{\textrm{min}}^{b} b L_{\textrm{min}}^{-b-1}, 
\]
or, equivalently $\frac{\alpha}{\alpha+1} < b$. With $b>1$, this always holds meaning that when the loads are Pareto distributed, there will always 
be an abrupt first order rupture at the attack size $p^{\star}=1-\frac{\bE{L}}{\bE{L}+\alpha L_{\textrm{min}}}=1-\frac{1}{1+\alpha\frac{b-1}{b}}$. In fact, we can see that this attack will lead to 
a complete breakdown of the system since for $x\geq L_{\textrm{min}}$, we have 
\begin{align}
&\frac{d}{dx} \left(\bP{L > x} \left(\alpha x + \bE{L ~|~ L > x} \right) \right)  
\\ \nonumber
& = \alpha \bP{L > x} - x p_L(x) (\alpha + 1)
\\ \nonumber
& = \alpha {L_{\textrm{min}}^b}{x^{-b}}- (\alpha + 1)xL_{\textrm{min}}^{b} b x^{-b-1}
\\ \nonumber
& = {L_{\textrm{min}}^b}{x^{-b}} \left(\alpha-b(\alpha+1)\right)
\\ \nonumber
& < 0
\end{align}
for any $\alpha>0$ and $b>1$. Therefore, it is always the case that $\left(\bP{L > x} \left(\alpha x + \bE{L ~|~ L > x} \right) \right)$ has a unique maximum 
at $x=L_{\textrm{min}}$, and the abrupt first order rupture completely breaks down the system.

These results show that for a given $L_{\textrm{min}}$ and $\bE{L}$ with $\bE{L} > L_{\textrm{min}}$, Pareto distribution is the {\em worst} possible scenario in terms of the overall robustness of the 
power system. Put differently, with $L_{\textrm{min}}$ and $\bE{L}$ fixed, the robustness curve $n_{\infty}(p)$ for the Pareto distribution constitutes a lower bound 
for that of any other distribution.
From a design perspective, we see that changing the tolerance parameter $\alpha$ will not help in mitigating the abruptness of the breakdown of the system in the case
of Pareto distributed loads. On the other hand, the point at which the abrupt failure takes place, i.e., the critical attack size $p^{\star}$ can be increased by increasing $\alpha$.

\subsection{Weibull distribution}
The last distribution we will consider is Weibull distribution, which has the form
\[
p_L(x) = \frac{k}{\lambda} \left(\frac{x-L_{\textrm{min}}}{\lambda} \right)^{k-1} e^{-\left(\frac{x-L_{\textrm{min}}}{\lambda} \right)^{k}}
 \1{x \geq L_{\textrm{min}}},
 \]
with $\lambda, k>0$.
The case $k=1$ corresponds to the exponential distribution, and $k=2$ corresponds to Rayleigh distribution. The mean load is given by 
$\bE{L}=L_{\textrm{min}}+ \lambda \Gamma (1+1/k)$, where $\Gamma(\cdot)$ is the gamma-function. 
As usual, we check the derivative of   $\left(\bP{L > x} \left(\alpha x + \bE{L ~|~ L > x} \right) \right)$ for $x \geq L_{\textrm{min}}$. 
On that range, we have $\bP{L > x} =  e^{-\left(\frac{x-L_{\textrm{min}}}{\lambda} \right)^{k}}$ so that
\begin{align}
\nonumber
&\frac{d}{dx} \left(\bP{L > x} \left(\alpha x + \bE{L ~|~ L > x} \right) \right) 
\\ 
& =  e^{-\left(\frac{x-L_{\textrm{min}}}{\lambda} \right)^{k}} \left(\alpha- (\alpha + 1)x \frac{k}{\lambda} \left(\frac{x-L_{\textrm{min}}}{\lambda} \right)^{k-1}\right)
\label{eq:derivative_of_weibull}
\end{align}
which becomes zero if
\begin{equation}
x (x-L_{\textrm{min}})^{k-1} = \frac{\alpha \lambda^{k}}{(\alpha+1)k}.
\label{eq:weibull_derivative_zero}
\end{equation}
This already prompts us to consider the cases $k<1$ and $k > 1$ separately. In fact, with $k > 1$, we see that 
$p_L(L_{\textrm{min}})=0$ and (\ref{eq:rupture_condition}) does {\em not} hold regardless of $\alpha$. In addition, 
there is one and only one $x>L_{\textrm{min}}$ that can satisfy (\ref{eq:weibull_derivative_zero}). Consequently,
 for $k \geq 1$ the system will always undergo a second-order transition with a diverging rate of failure before 
 breaking down completely through a first-order transition. 
 
  \begin{figure}[!t]
\centering{
\hspace{-0.5cm} 
\includegraphics[totalheight=0.3\textheight]{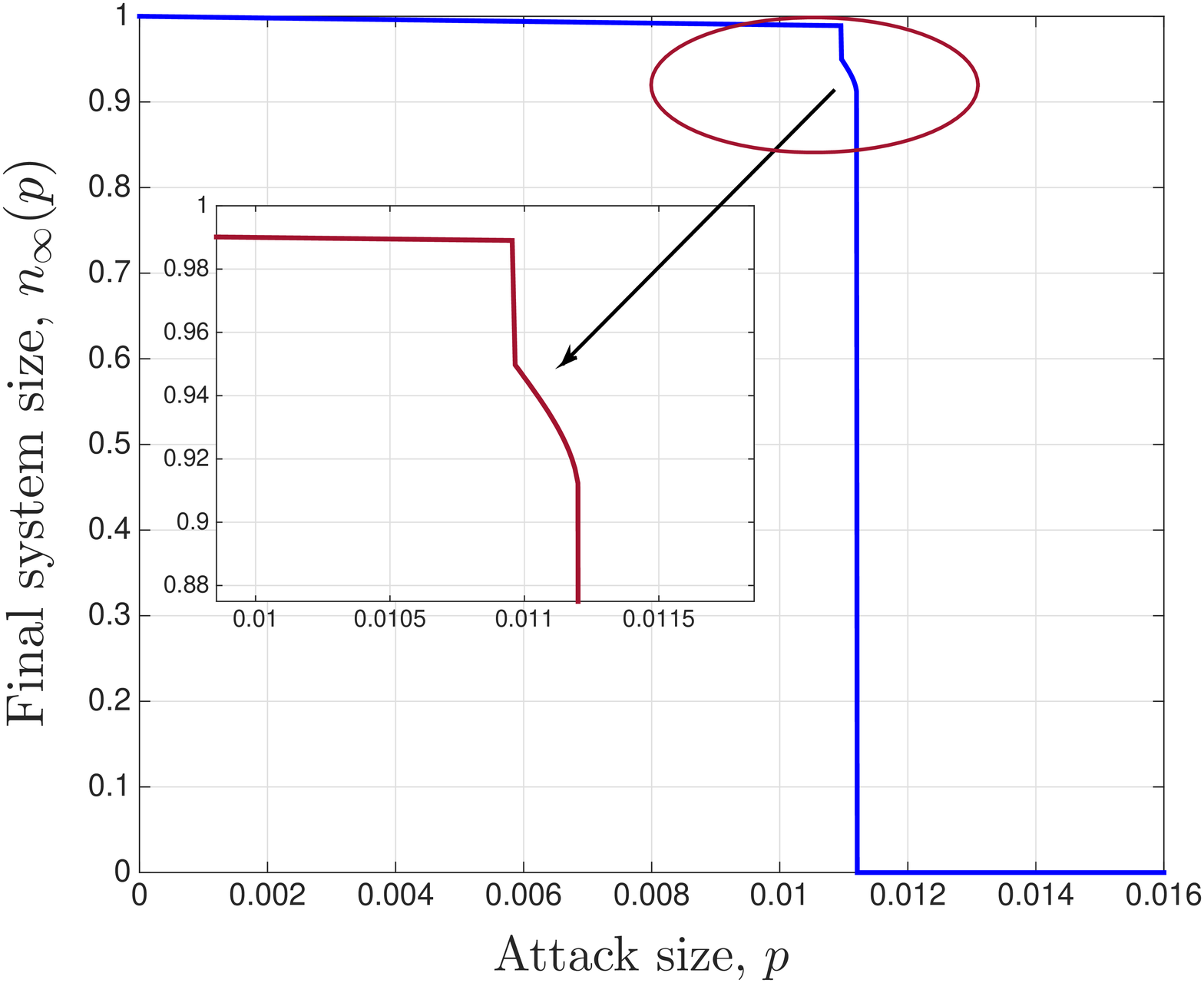}}
\caption{\sl The two stage breakdown of the system is demonstrated, where $L_1, \ldots, L_N$ are drawn 
from Weibull distribution with $k=0.8$, $\lambda=150$, $L_{\textrm{min}}=10$, $\alpha=0.2$. We plot the relative final size
$n_{\infty}(p)$ as a function of the attack size $p$. The Inset zooms in to the region where the system goes through a series of 
first-order, second-order, and then again a first-order transition.}
\label{fig:2_step} 
\end{figure}

The case $k<1$ gives an entirely different picture since $p_L(L_{\textrm{min}})=\infty$ 
and (\ref{eq:rupture_condition})  {\em always} holds regardless of $\alpha$. So, the system will always go through an
abrupt first order transition at the attack size $p^{\star}=1-\frac{\bE{L}}{\bE{L}+\alpha L_{\textrm{min}}}$. Whether this rupture will entirely breakdown the system 
depends on the existence of the solutions of (\ref{eq:weibull_derivative_zero}). It is easy to see that $x (x-L_{\textrm{min}})^{k-1}$ takes its minimum value at $x=L_{\textrm{min}}/k$
and equals to
$\frac{L_{\textrm{min}}^{k}}{k} \left(\frac{1-k}{k}\right)^{k-1}$. Thus, if it holds that
\begin{equation}
L_{\textrm{min}}^{k} \left(\frac{1-k}{k}\right)^{k-1} > \frac{\alpha \lambda^{k}}{\alpha+1},
\label{eq:potential_equality}
\end{equation}
then (\ref{eq:weibull_derivative_zero}) has no solution and the derivative given
at (\ref{eq:derivative_of_weibull}) is negative for all $x \geq L_{\textrm{min}}$, meaning that
$\bP{L > x} \left(\alpha x + \bE{L ~|~ L > x} \right)$ is maximized at $x=L_{\textrm{min}}$.
Then, the abrupt first order rupture at 
 $p^{\star}=1-\frac{\bE{L}}{\bE{L}+\alpha L_{\textrm{min}}}$ will indeed breakdown the system completely. The same conclusion follows
 if (\ref{eq:potential_equality}) holds with equality by virtue of the fact  that (\ref{eq:derivative_of_weibull}) is again non-positive for all $x \geq L_{\textrm{min}}$.

On the other hand, if 
 \begin{equation} 
 L_{\textrm{min}}^{k} \left(\frac{1-k}{k}\right)^{k-1} < \frac{\alpha \lambda^{k}}{\alpha+1},
 \label{eq:two_step_transition}
\end{equation}
then (\ref{eq:weibull_derivative_zero}) will have two solutions both with $x > L_{\textrm{min}}$. 
This implies that $\bP{L > x} \left(\alpha x + \bE{L ~|~ L > x} \right)$ has another maximum at a point $x > L_{\textrm{min}}$.
If this maximum is indeed the global maximum (i.e., it is larger than the maximum attained at $x = L_{\textrm{min}}$),
then the system will go under two first-order phase transitions before breaking down.
First, an abrupt rupture will take place at $p^{\ast}=1-\frac{\bE{L}}{\bE{L}+\alpha L_{\textrm{min}}}$. But, this won't break down the system completely and $n_{\infty}(p^{\ast^{+}})$ will be positive.
As $p$ increases further, we will observe a second-order transition with a diverging rate of failure until another first-order rupture breaks down the system completely. 
We demonstrate this phenomenon in Figure \ref{fig:2_step}, where we set
$k=0.8$, $\lambda=150$, $L_{\textrm{min}}=10$, $\alpha=0.2$. We emphasize that this behavior (i.e., occurrence of two first-order transitions) is not immediately warranted under (\ref{eq:two_step_transition}).
It is also needed that $\bP{L > x} \left(\alpha x + \bE{L ~|~ L > x} \right)$ has a global maximum at a point $x > L_{\textrm{min}}$.

\section{Numerical results}
\label{sec:Simu}

We now check the validity of our 
mean-field analysis for finite number $N$ of power lines
via simulations. We will do so with an eye towards comparing
the robustness of power systems under different distributions of loads.

 \begin{figure}[!t]
\centering{
\hspace{-0.5cm} 
\includegraphics[totalheight=0.3\textheight]{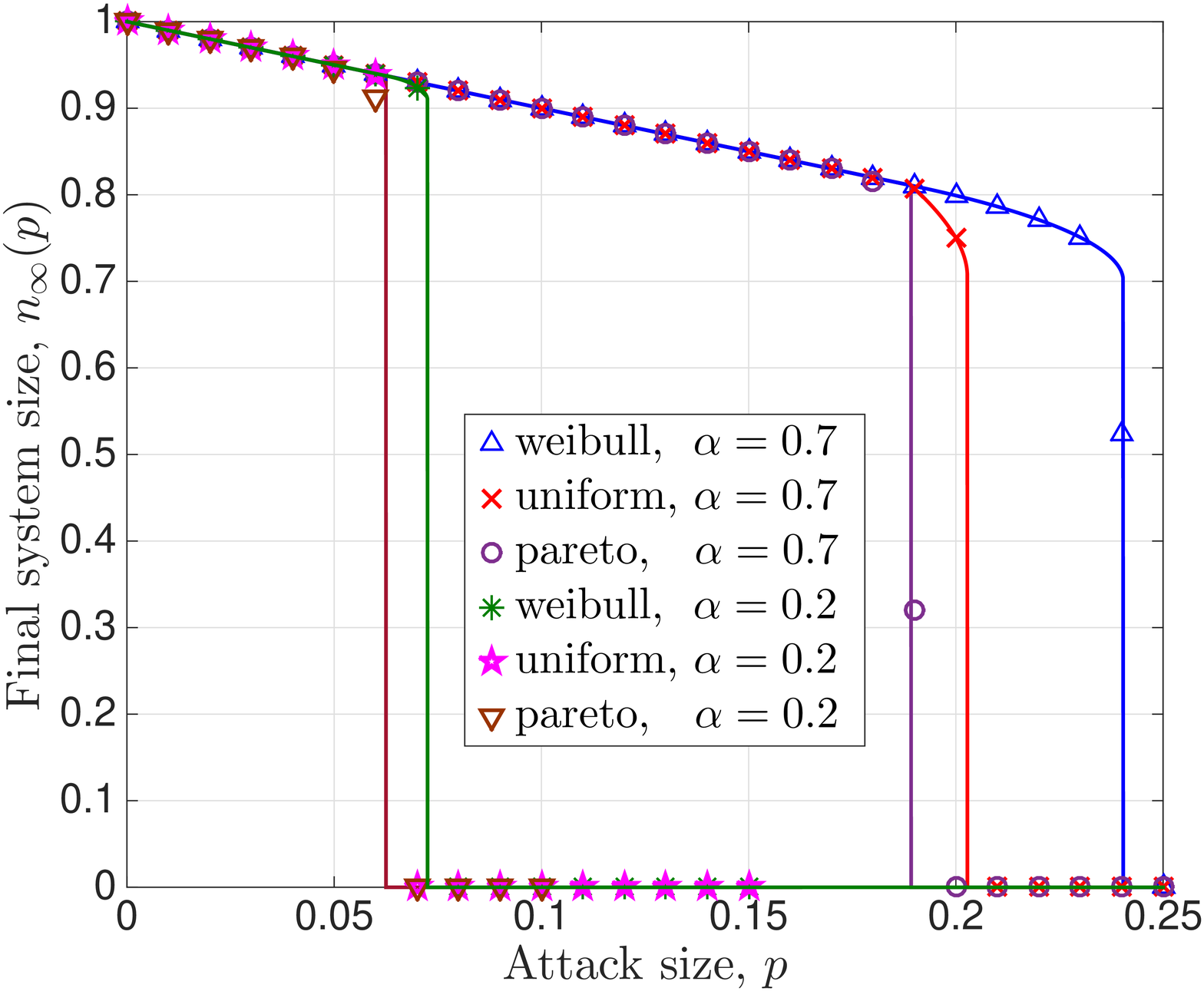}}
\caption{\sl (Color Online). We plot $n_{\infty}(p)$ vs.~$p$ under six different cases. Analytical results are represented by lines, whereas empirical results (obtained through averaging over 500 independent runs) 
are represented by symbols. 
We set $N=100,000$, $L_{\textrm{min}}=10$, and $\bE{L}=30$. For the case when $L_1,\ldots, L_N$
follow a Weibull distribution, we take the shape parameter to be $k=2$, leading to 
a scale parameter $\lambda=22.5676$. We see that numerical results match 
 the analytical results very well. }
\label{fig:numerical} 
\end{figure}

In the first batch of simulations, we fix the minimum load at $L_{\textrm{min}}=10$ and mean
load at $\bE{L}=30$. These constraints fully determine the load distribution $p_{L}(x)$ in the cases where $p_L$ is 
Uniform (with $L_{\textrm{min}}=10$ and $L_{\textrm{max}}=50$) or Pareto (with $L_{\textrm{min}}=10$ and $b=1.5$). For the case
where $p_{L}$ is Weibull, we need to pick $k$ and $\lambda$ such that $\lambda \Gamma(1+1/k)=20$, where $\Gamma(x)=\int_{0}^{\infty} t^{x-1} e^{-t} dt$. 
We consider $k=2,\lambda=22.5676$ as an example point. 

Our simulation set up is as follows. We fix the number of lines $N$, and generate $N$ random variables from the given distribution $p_L(x)$ corresponding to loads
$L_1,\ldots,L_N$. Then, for a given $p$, we perform an attack on $\lceil pN \rceil$ lines that are selected uniformly at random and assume those lines have failed. Next, using the democratic redistribution 
of loads, we iteratively fail any line whose load exceeds its capacity (which is set to $(1+\alpha)$ times its initial load). We consider two possible tolerance
parameters: i) $\alpha=0.2$ and ii) $\alpha=0.7$. 
The process stops when the system is stable; i.e., all lines have a load below their capacity.
Of course, this steady state can be reached at a point where all lines in the system have failed. We record the corresponding fraction of active lines at the steady state. This process is repeated independently
500 times for each $p$, and the average fraction of active lines at the steady state over 500 independent runs gives the empirical value of $n_{\infty}(p)$. We then compare this with the quantity obtained from our analysis
in Section \ref{sec:Results}. 

Results are depicted in Figure \ref{fig:numerical}. First of all, we see an almost perfect agreement between our mean-field analysis and numerical results. 
It is worth noting that the fit between analysis and simulations required particularly large values of $N$ in the case of Pareto distribution. For the other two distributions, even $N=5000$ leads to almost 
perfect agreement. Focusing on two curves corresponding to uniform distribution, we see from Figure \ref{fig:numerical} that the tolerance parameter $\alpha$ not only changes the maximum attack size $p^{\star}$ that the system can sustain (in the sense of not breaking down entirely), but it can also affect the type of the phase transition. In particular, with $\alpha=0.2$, an abrupt failure takes place at $p^{\star}=0.0625$, whereas with
$\alpha=0.7$ the system goes through a second order transition starting with the attack size $p=0.189$, and then breaks down entirely through a first-order jump at $p^{\star}=0.203$. 

As expected from our previous
discussion, the distribution that leads to the worst robustness is Pareto among all distributions considered here. However, we see that under certain conditions uniform and Weibull distributions can match the poor robustness characteristics of the Pareto distribution; one example is the case shown in Figure \ref{fig:numerical} with uniform load distribution and  $\alpha=0.2$. Finally, we observe that Weibull distribution can lead to a 
significantly better robustness than Pareto and Uniform distribution, under the same mean and minimum load. 

\begin{figure}[t]
\centering{
\hspace{-0.5cm} 
\includegraphics[totalheight=0.3\textheight]{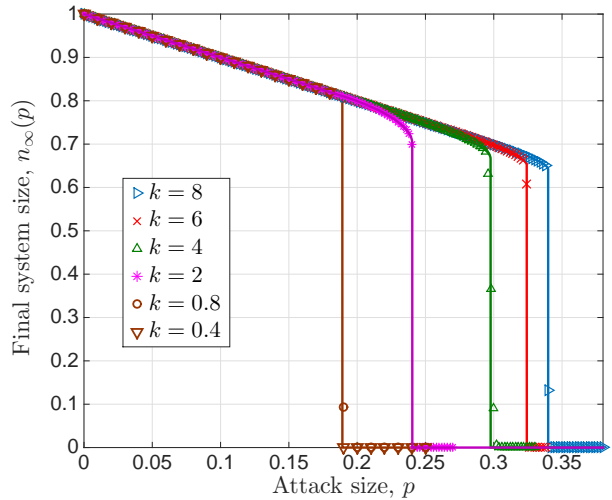}}
\caption{\sl (Color Online). We plot $n_{\infty}(p)$ vs.~$p$ when 
$L_1,\ldots, L_N$
follow a Weibull distribution
 with $L_{\textrm{min}}=10$, $\bE{L}=30$. We set $N=100,000$ and $\alpha=0.7$.
Analytical results are represented by lines, whereas empirical results (obtained through averaging over 500 independent runs) are represented by symbols. 
Again, we see that numerical results match  the analytical results pretty well. }
\label{fig:weibull_comp} 
\end{figure}

The last observation worths investigating further. In particular, even when $L_{\textrm{min}}$ and $\bE{L}$ are fixed, the Weibull
distribution has another degree of freedom; i.e., parameters $k$ and $\lambda$ are arbitrary subject to the condition
that $\lambda \Gamma(1+1/k) = \bE{L}$. In order to understand the effect of the shape parameter $k$ in the robustness of power systems under Weibull distributed loads, 
we ran another set of simulations with $N=100,000$, $\alpha=0.7$,  $L_{\textrm{min}}=10$, $\bE{L}=30$, and $\lambda=\frac{20}{\Gamma(1+1/k)}$ for various values of $k$.
Results are depicted in Figure \ref{fig:weibull_comp} where again analytical results are represented by lines
and empirical results (obtained through averaging over 500 independent runs) are represented by symbols. We again observe an excellent match between analytical 
and numerical results. We remark that with the given parameter setting, the cases where $k \leq 1$ all result in the same robustness behavior with an abrupt first-order
rupture at $p=0.189$.

More importantly, we see that the robustness of the system improves as the parameter $k$ increases. 
It is known that as $k$ gets larger the Weibull distribution gets closer and closer to a Dirac delta distribution centered
at its mean. In other words, as $k$ goes to infinity the Weibull distribution converges to a degenerate distribution and loads
$L_1,\ldots, L_N$ will all be equal to the mean $\bE{L}$. This naturally prompts us to ask whether a degenerate distribution of loads is the 
universally optimum strategy among all possible distributions with the same mean $\bE{L}$, with optimality criterion being the maximization
of robustness against random attacks or failures. Here, a natural condition 
for maximization of robustness would be to maximize the critical attack size $p^{\star}$.
We answer this question, in the affirmative, in the next section.

\section{Optimal load distribution}
\label{sec:optimal}
To drive the above point further 
and to better understand the impact of the shape parameter $k$ on the system robustness,
we now plot the maximum attack size $p^{\star}$ as a function of $k$ under the same setting;
 see Figure \ref{fig:varying_k}. Namely, we let $L_1,\ldots, L_N$
follow a Weibull distribution
 with $L_{\textrm{min}}=10$, and $\lambda=20/\Gamma(1+1/k)$ so that $\bE{L}=30$.
 We see from Figure \ref{fig:varying_k} that, in all choices of $\alpha$ considered here, 
 the maximum attack size $p^{\star}$
 is monotone increasing with $k$; note that $p^{\star}$ is  seen to be constant over the range $0<k\leq 1$. 
 It is also evident from Figure \ref{fig:varying_k}  that $p^{\star}$ tends to {\em converge} to a fixed value as $k \to \infty$. On the other hand,
 with $k \to \infty$, we know that Weibull distribution converges to a Dirac delta distribution centered at $\bE{L}$. 
It is therefore of interest to check whether $p^{\star}$ is always maximized by choosing all loads $L_1, \ldots, L_N$ equally, i.e., by choosing $p_L(x)$
to be a degenerate distribution with mean $\bE{L}$ and zero variance. 
 
 \begin{figure}[t]
\centering{
\hspace{-0.5cm} 
\includegraphics[totalheight=0.3\textheight]{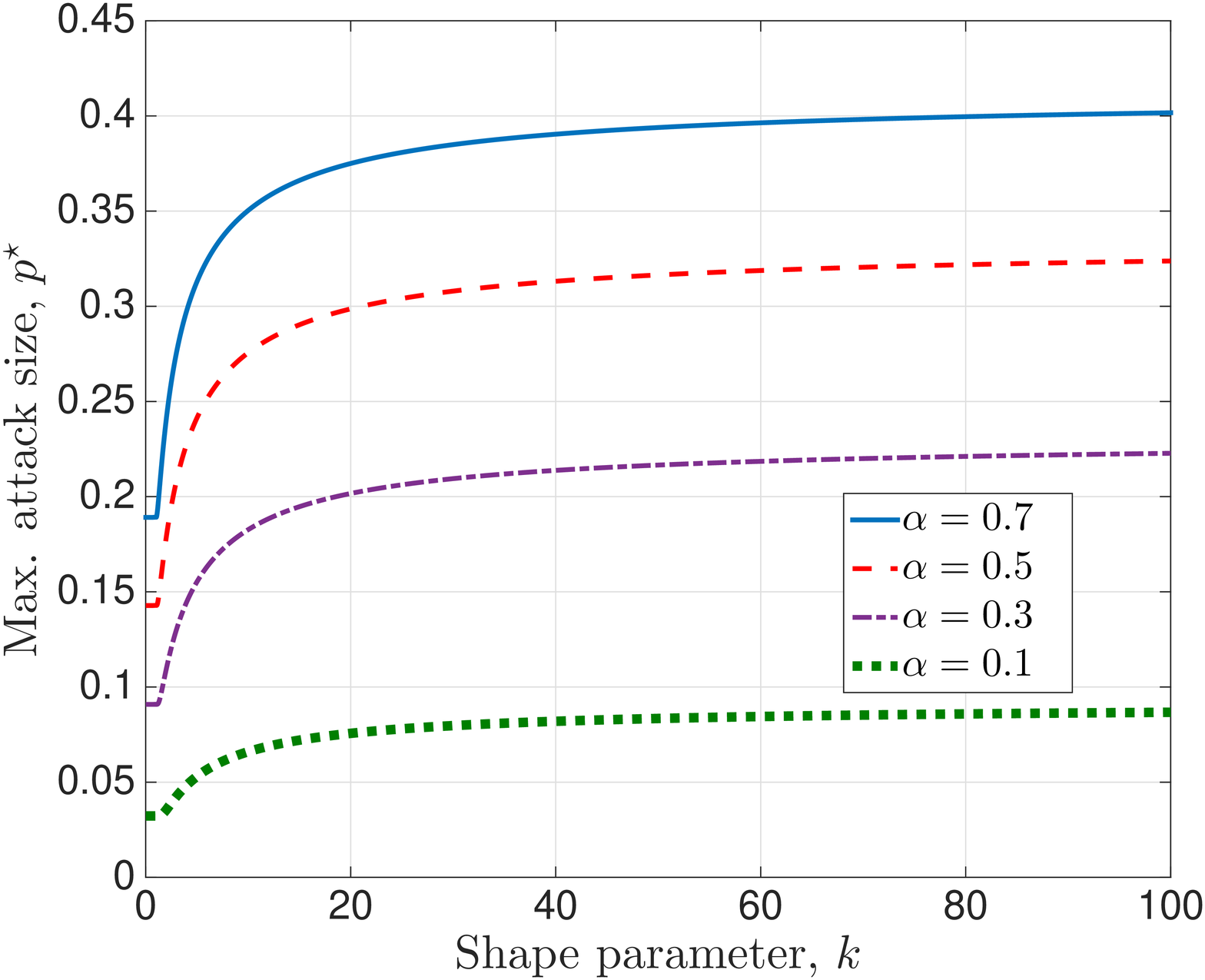}}
\caption{\sl (Color Online). We plot the maximum attack size $p^{\star}$, when 
$L_1,\ldots, L_N$
follow a Weibull distribution
 with $L_{\textrm{min}}=10$, $\bE{L}=30$, as a function of the shape parameter $k$ of the Weibull distribution. 
 We set $N=100,000$ and consider four tolerance parameters $\alpha=0.1, 0.3, 0.5, 0.7$.
The curves correspond to analytical results computed directly from (\ref{eq:max_attack}).
}
\label{fig:varying_k} 
\end{figure}

Let $p_L(x)$ be an arbitrary distribution with mean $\bE{L}$, and assume that $p_L(x) =0$ for $x \leq 0$; i.e., that $L$
is non-negative. Recall that maximum attack size $p^{\star}$ is given by (\ref{eq:max_attack}) and observe that
\begin{align}
& \bP{L > x} \left(\alpha x + \bE{L ~|~ L > x} \right)
\nonumber \\
& = \alpha x \bP{L > x} + \bE{L \cdot \1{L>x}}  
\nonumber \\
& \leq \alpha \bE{L} +  \bE{L \cdot \1{L>x}}  
\label{eq:optimality_step1} \\
& \leq (\alpha+1) \bE{L}, 
\label{eq:optimality_step2}
\end{align}
for any $x \geq 0$. In (\ref{eq:optimality_step1}) we used the {\em Markov Inequality} \cite[p. 151]{papoulis2002probability}, i.e., the fact that $\bP{L>x} \leq \bE{L}/x$
for any non-negative random variable $L$ and $x \geq 0$. Reporting (\ref{eq:optimality_step2}) into  (\ref{eq:max_attack}), we get
\begin{equation}
p^{\star} \leq 1- \frac{\bE{L}}{(\alpha+1) \bE{L}}=\frac{\alpha}{\alpha+1}.
\label{eq:max_attack_upper}
\end{equation}

This shows that the maximum attack size can never exceed $\frac{\alpha}{\alpha+1}$ under any choice of load distribution. 
On the other hand, consider the case where $p_L(x) = \delta(\bE{L})$ with $\delta(\cdot)$ denoting a Dirac delta function. This implies 
that $L_1= \cdots = L_N = \bE{L}$. Let $p^{\star}_{\textrm{dirac}}$ denote the corresponding
maximum attack size.
With $x=\bE{L}^{-}$, we have $\bP{L > x}=1$ and  $\bE{L \cdot \1{L>x}}  =\bE{L}$. Thus,
\begin{align}
\lim_{x\uparrow \bE{L}} \alpha x \bP{L > x} + \bE{L \cdot \1{L>x}}  = (\alpha+1) \bE{L}
\nonumber
\end{align}
so that 
\[
\max_x\{\bP{L > x} \left(\alpha x + \bE{L ~|~ L > x} \right)\} \geq  (\alpha+1) \bE{L}.
\]
Invoking  (\ref{eq:max_attack}), this leads
\[
p^{\star}_{\textrm{dirac}} \geq \frac{\alpha}{\alpha+1}.
\]
But, (\ref{eq:max_attack_upper}) holds for any distribution and hence is also valid for $p^{\star}_{\textrm{dirac}}$. 
Combining these, we obtain that
\begin{align}
p^{\star}_{\textrm{dirac}} = \frac{\alpha}{\alpha+1}.
\label{eq:optimal_dirac}
\end{align}
This establishes that a degenerate distribution is indeed {\em optimal} for any given mean value of the load, and the achieved maximum attack size
is given by $\alpha/(\alpha+1)$. What is even more remarkable is that, this maximum attack size is independent of the mean load $\bE{L}$.

It is now clear to what point the curves in Figure \ref{fig:varying_k} tend to converge as $k \to \infty$; they can indeed be seen to get closer and closer to the corresponding value of $\alpha/(\alpha+1)$.
We close by demonstrating the variation of the final system size as a function of the attack size, in the case where loads follow a Dirac distribution. We easily see that 
$\bP{L > x} \left(\alpha x + \bE{L ~|~ L > x} \right)$ increases linearly for $x<\bE{L}$ and equals to zero for $x \geq \bE{L}$. Therefore, the breakdown of the system will always be through an abrupt first order
rupture. 

This is demonstrated in Figure \ref{fig:dirac}, where it is seen once again that numerical results match the analysis perfectly.
Comparing these plots with Figures \ref{fig:numerical} and \ref{fig:weibull_comp}, we see the dramatic impact that the load distribution 
has on the robustness of a power system. For instance, with $\alpha=0.2$ and mean load fixed at $30$ we see that maximum attack size that the system 
can sustain is 6.3\% for Pareto and Uniform distributions whereas it is 17\% when all loads are equal. Similarly, with $\alpha=0.7$ we see that maximum 
attack size is  18\% for Pareto distribution and 19\% for Uniform distribution, while for the Dirac delta distribution, it increases to 41\%. 
These findings suggest that under the democratic fiber bundle-like model considered here, power systems with homogenous loads are significantly 
more robust against random attacks and failures, as compared to systems with heterogeneous load distribution.

 \begin{figure}[t]
\centering{
\hspace{-0.5cm} 
\includegraphics[totalheight=0.3\textheight]{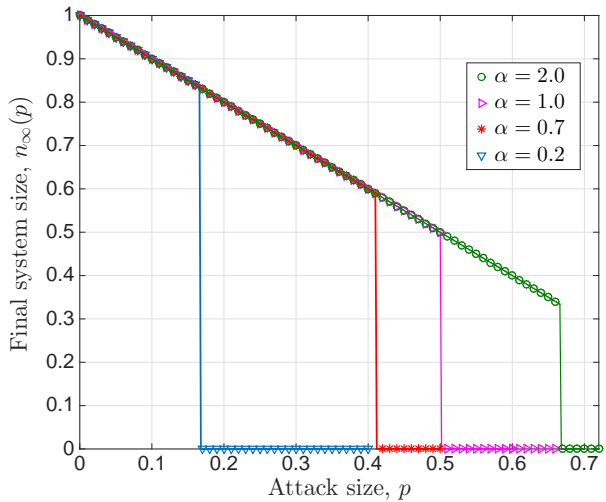}}
\caption{\sl (Color Online). We plot the final system size $n_{\infty}(p)$ as a function of the attack size $p$,
when $L_1= \cdots = L_N =\bE{L}$. For the numerical results, we take $N=100,000$, $\bE{L}=30$,
and consider four tolerance parameters $\alpha=0.2, 0.7, 1.0, 2.0$. Each data point (represented by a symbol) 
is the result of averaging over 500 independent runs.
The lines correspond to analytical results computed directly from (\ref{eq:max_attack}).  We see a perfect agreement between analysis
and experiments. In all cases, the system breakdowns abruptly through a first order transition at $p^{\star}=\frac{\alpha}{\alpha+1}.$}
\label{fig:dirac} 
\end{figure}
  
\section{Conclusion}
\label{sec:Conclusion}

We studied the robustness of power systems consisting of $N$ lines
under a democratic-fiber-bundle like model and against random attacks. We show that the system 
goes under a total breakdown through a first-order transition as the attack size reaches a critical value.
We derive the conditions under which the first-order rupture occurs abruptly without any 
preceding divergence of the failure rate; those situations correspond to cases where no cascade of failures occurs until a critical attack
size is reached, followed by a total breakdown at the critical attack size. Numerical results 
are presented and confirm the analytical findings. Last but not least, we prove that with mean load fixed,
robustness of the power system is maximized when the variation among the line loads is minimized. In other words,
 a Dirac delta load distribution leads to the optimum robustness.

Our results highlight how different parameters of the load distribution and the power line capacity 
affect the robustness of the power grid against failures and attacks. To that end,
our results can help derive guidelines for the robust design of the power grid. 
We believe that the results presented here give very interesting insights 
into the cascade processes in power grids, although through a very simplified model of the grid.
The obtained results can be useful in other fields as well, where equal redistribution of flows is a reasonable assumption. 
Examples include traffic jams, landslides, etc.

There are many open problems one can consider for future work. 
For instance, the analysis can be extended to the case where the tolerance
parameter $\alpha$ is not the same for all lines, but follows a given probability distribution. 
It would be interesting to see if the robustness is still maximized with 
a {\em narrow} distribution of $\alpha$. It may also be of interest to 
study robustness against {\em targeted} attacks rather than random failures.

\section*{Acknowledgments}
This research was supported in part by National Science Foundation through grant CCF \#1422165, and by
the Department of Electrical and Computer Engineering at Carnegie Mellon University.

\bibliography{references}

\end{document}